Investigating viscous surface wave propagation modes and study of nonlinearities in a finite depth fluid

Arash Ghahraman[a] *, Gyula Bene [b] **
[a] Doctoral School of Physics, Institute of Physics, Eötvös Loránd University, , H-1117 Budapest, Pázmány Péter sétány 1/A
[b] Department of Theoretical Physics, Institute of Physics, Eötvös Loránd University, H-1117 Budapest, Pázmány Péter sétány 1/A
*arash.ghgood@gmail.com
** bene@arpad.elte.hu

Abstract
The object of this study is to investigate the effect of viscosity on the propagation of free-surface waves in an incompressible viscous fluid layer of arbitrary depth. While we provide a more detailed study of properties of linear surface waves, the description of fully nonlinear waves in terms of KdV-like equations is discussed. In the linear case, we find that in shallow enough fluids, no surface waves can propagate. Even in any thicker fluid layers, propagation of very short and very long waves is forbidden. When wave propagation is possible, only a single propagating mode exists for any given horizontal wave number. The numerical results show that there can be two types of non-propagating modes. One type is always present, and there exist still infinitely many of such modes at the same parameters. In contrast, there can be zero, one or two modes belonging to the other type. Another significant feature is that KdV-like equations. They describe propagating nonlinear viscous surface waves. Since viscosity gives rise to a new wavenumber that cannot be small at the same time as the original one, these equations may not exist. Nonetheless, we propose a reasonable nonlinear description in terms of 1+1 variate functions that make possible successive approximations.

## 1. Introduction

The propagation of water waves over a fluid is a long run issue both of mathematics, fluid mechanics, hydrogeology, coastal engineering, etc. [1, 2]. Historically, much research has been devoted to inviscid fluid the scope has been broadened with time. Various equations have been proposed to model this propagation of water waves. The goal is to find reduced models, particularly in dimentionality with as little number of fields as possible, should they be valid only in an asymptotic regime [1]. However, even if in most situations in coastal engineering, the assumption of inviscid flow leads to very accurate results, there are other physical scenarios where the viscosity needs to be taken into account [3, 4, 5]. Moreover, there are many situations in which the viscosity and surface tension should be considered [2, 6, 7].

In the inviscid case, the complete and precise explanation of most water waves models has been recently accomplished [8, 1]. Things are more delicate when viscosity is taken into account, and complete justification of the asymptotic models is still lacking [1]. Therefore, among the different free surface problems, those involving viscous flows are the most difficult ones [9, 10, 11]. In addition to the nonlinear boundary conditions at the free surface, the flow field is given by the nonlinear Navier-Stokes equations.

Boussinesq [12] and Lamb [13] studied the effect of viscosity on free surface waves. They concentrated on linearized NS equations on deep-water and computed the dispersion relation. Basset [14] also worked on viscous damping of water waves. The first formal dissipative KdV equation presented by Ott and Sudan [15]. Furthermore, Longuet-Higgins [16, 17] found Lamb's coefficient by using a boundary layer model and inspecting the equivalence of this model with the theory on weakly-damped waves by Ruvinsky & Freiman [18]. Lighthill [19] used the deep-water inviscid solution to approximate the dissipative terms inside the kinetic energy equation for a viscous fluid. In other words, over the years, the deep-water conditions have been used in several works that confirmed the validity of Lamb's result through different techniques and approaches. By assuming small amplitude gravity waves in a constant depth of water, Hunt studied the rate of attenuation of wave amplitude in shallow water [3]. Although these results are valid for an arbitrary depth, surface tension is not considered in his studies.

To construct an accurate viscous model, one has to do first a linear study. Because of the various dimensionless parameters, it is necessary to determine the most engaging regime between the parameters [1]. Also, one must either assume linearized Navier-Stokes (NS) equations or justify that the nonlinear terms can be dropped. In 1975, Kakutani and Matsuuchi [20] started from the NS equations and performed a clean boundary layer analysis. First, they make a linear analysis that gives the dispersion relation distinguish various regimes of Re as a function of the small classical parameter of any Boussinesq study. Then, they derive the corresponding viscous KdV equation [1]. Matsuuchi tried to validate the equation that they were led to. He shows that their model does not modify the number of crests found by an experimental study and by a (non-viscous) KdV simulation, but induces a shift in phase [1, 21]. Indeed, the regime is not the Boussinesq one. The most plausible formulation of the KdV equation modified by viscosity was given by Miles [22]. He includes the effect of both dispersion and dissipation. Miles was motivated by his study of the evolution of a solitary wave for very weak non-linearity. In 1987, Khabakhpashev [23] extended the derivation of the viscous KdV evolution equation to the derivation of a Boussinesq system, studied the dispersion relation and predicted a reverse flow in the bottom in case of the propagation of a soliton wave. The proposed viscous KdV equation contains, however, an integral term which despite of the concise form makes the equations equivalent to an infinite set of differential equations.

Between 2004 and 2012, researchers concentrate on linearized Navier-Stokes (NS) equations. The most significant papers were published by Dias *et al.* in 2008 [24, 25]. More precisely, the authors take the (linearized) NS equations of deep-water flow with a free boundary. In this case, they use the Helmholtz decomposition, and both Bernoulli's equation (through an irrotational pressure) and the kinematic boundary are modified. At the end of their paper, they also discuss nonlinearities and *ad hoc* modeling. Starting from such a model in the deep water case, they end up at the dissipative generalization of the Non-Linear Schrödinger (NLS). However, surface tension is not considered in their studies. In addition, in many cases in nonlinear wave dynamics, such as for internal solitary waves in the ocean, the nonlinearity is not so weak as is implied by the KdV equation. Therefore a higher-order KdV equation can be obtained by considering the next order of the perturbation theory. The contribution of the various high-order terms depends on the parameters of the model such as shear flow, and sometimes several of them may be more important. In the presence of damping the modified KdV equation is not fully integrable. This historical point of view shows that although many studies have been done for linear surface water waves, few studies have been done for KdV-type equations and nonlinear surface water waves propagating in a viscous fluid. In this way, we aim to investigate the existence of soliton solutions and their decay to the damped KdV-type equation as well as study the effect of the viscosity on the wavelengths and the propagation of free-surface waves in an incompressible viscous fluid of arbitrary depth. We expand this situation with and without surface tension in details. Surprisingly, we obtain exciting results, particularly in the linear case, which are not mentioned in other works of literature. This article outlines the derivation of a system PDEs for surface water waves in the KdV-type regime, taking into account the viscosity and surface tension. For this purpose, we apply a procedure for finding the Taylor coefficients of a velocity field expanded in terms of the vertical coordinates at a boundary point [26]. So, velocity models that are polynomials in terms of the distance from the bottom are constructed.

The paper is organized as follows. In section 2., a derivation of the dispersion relation is given and discussed. The result underlies our further investigations. In section 3., different modes are discussed. Section 4. is devoted to a study of parameter dependence. Sections 2-4 refer to the linear case. The nonlinear situation is studied in section 6. The results are discussed in the concluding section 6.

## 2. Dispersion relation

We consider linear surface waves on a viscous, incompressible fluid of finite depth $h$. The coordinates $x$ and $y$ are horizontal, $z$ is vertical. The origin lies at the undisturbed fluid surface. Suppose that the flow corresponding to the surface wave does not depend on $y$. We start with the Ansatz

$$u = f' \exp(i\varphi) \tag{1}$$

$$w = -ik\, f \exp(i\varphi) \tag{2}$$

for the horizontal $u$ and vertical $w$ velocity components, where the complex valued function $f$ depends on $z$ only, the prime denotes derivative with respect to its argument, and phase $\varphi$ is given by

$$\varphi = kx - \omega t \tag{3}$$

Here $k$ is a real, positive wave number, while ω is in general complex, its imaginary part describing the damping. Note that the Ansatz automatically satisfies the incompressibility condition $\nabla \mathrm{v} = 0$. The linearized Navier-Stokes equation may be written as

$$\frac{\partial \mathbf{v}}{\partial t} - \nu \,\Delta \mathbf{v} = \nabla\left( -\frac{p}{\rho} - gz \right) \tag{4}$$

Since the right hand side is a full gradient, we have

$$\frac{\partial}{\partial z}\left( \frac{\partial u}{\partial t} - \nu \,\Delta u \right) = \frac{\partial}{\partial x}\left( \frac{\partial w}{\partial t} - \nu \,\Delta w \right) \tag{5}$$

Putting Ansatz (1) and (2) into Eq. (5), we obtain

$$f'''' + \left( i\frac{\omega}{\nu} - 2k^2 \right) f'' - \left( i\frac{\omega}{\nu} - k^2 \right) k^2 f = 0 \tag{6}$$

Eq. (6) has exponential solutions $f = \exp(mz)$. For the exponent $\kappa$ we get

$$m^4 + \left( i\frac{\omega}{\nu} - 2k^2 \right) m^2 - \left( i\frac{\omega}{\nu} - k^2 \right) k^2 f = 0 \tag{7}$$

The solutions are

$$m_{1,2} = \pm k \tag{8}$$

$$m_{3,4} = \pm\sqrt{k^2 - i\frac{\omega}{\nu}} \tag{9}$$

or brevity, we shall use the notation $k$ for $m_1 = -m_2$ and $\kappa$ for $m_3 = -m_4$. The general solution for $f$ may be given as

$$f = a_1 \cosh\left[k(z+h)\right] + a_2 \sinh\left[k(z+h)\right] + b_1 \cosh\left[\kappa(z+h)\right] + b_2 \sinh\left[\kappa(z+h)\right] \tag{10}$$

where $a_1$, $a_2$, $b_1$ and $b_2$ are integration constants and $h$ stands for the fluid depth. Then boundary conditions at the bottom,

$$u(z=-h) = w(z=-h) = 0 \tag{11}$$

Implies that

$$a_1 + b_1 = 0 \tag{12}$$

$$a_2 k + b_2 \kappa = 0 \tag{13}$$

So we have

$$a_1 = A,\; b_2 = B,\; a_2 = -\frac{\kappa}{k}B,\; b_1 = -A \tag{14}$$

expressed in terms of the new constants $A$ and $B$. Hence for $f$ we get

$$f = A\cosh\left[k(z+h)\right] - \frac{\kappa}{k}B\sinh\left[k(z+h)\right] - A\cosh\left[\kappa(z+h)\right] + B\sinh\left[\kappa(z+h)\right]. \tag{15}$$

Upon integrating the $x$ component of the Navier-Stokes equation with respect to $x$, we get the pressure as

$$p = p_0 - \rho g z - \rho \mathrm{e}^{(i\varphi)} \left( -\frac{\omega}{k} f' - i\nu k f' + i\frac{\nu}{k} f''' \right) . \tag{16}$$

At the fluid surface we have the boundary conditions that the strain forces are continuous, therefore in linear approximation, we have

$$u_z + w_x = 0 \tag{17}$$

for the shear and

$$p - 2\rho\nu\, w_z = p_0 - \sigma \eta_{xx} \tag{18}$$

for the pressure. Here $\eta = \eta(x,t)$ stands for the deviation of the fluid surface from equilibrium and $\sigma$ is the surface tension.
Eq. (17) implies

$$f'' + k^2 f = 0 \tag{19}$$

at $z = 0$, while Eq. (18) implies

$$-g\eta + \frac{\sigma}{\rho}\eta_{xx} - \mathrm{e}^{i\varphi}\left( -\frac{\omega}{k} f' - i\nu k\, f' + i\frac{\nu}{k} f''' \right) + 2i\nu k\, f' \mathrm{e}^{i\varphi} = 0 \ . \tag{20}$$

In the case of the linear approximation, the following equation is satisfied at the surface

$$\eta_t = w \tag{21}$$

Note that on the right hand side we may set $z = 0$. Inserting Eq. (2) and combining the result with Eq. (20), we have

$$\left(1 + \frac{\sigma}{g\rho}k^2\right)k^2 f - \frac{1}{g}\left(\omega^2 + 3i\omega\nu k^2\right) f' + i\frac{\omega\nu}{g} f''' = 0 \tag{22}$$

or, in terms of $\kappa$ in Eq. (9)

$$\left(1 + \frac{\sigma}{g\rho}k^2\right)k^2 f + \frac{\nu^2}{g}(\kappa^2 - k^2)(\kappa^2 + 2k^2) f' - \frac{\nu^2}{g}(\kappa^2 - k^2) f''' = 0 \tag{23}$$

Where $z = 0$. Substituting the solution (15) into (19) and (23) we obtain a linear homogeneous system of equations for the coefficients $A$ and $B$. The vanishing of the determinant of this system, which is the condition for the existence of a nontrivial solution, may be expressed in terms of the dimensionless variables

$$K = kh \qquad Q = \kappa h \tag{24}$$

$$p = \frac{\nu^2}{gh^3} \qquad s = \frac{\sigma}{\rho g h^2} \tag{25}$$

as

$$\begin{aligned} &K\left(Q\sinh K\cosh Q - K\cosh K\sinh Q\right)(1 + sK^2) + p\Big[-4K^2Q\left(K^2 + Q^2\right) \\ &+Q\left(Q^4 + 2K^2Q^2 + 5K^4\right)\cosh K\cosh Q - K\left(Q^4 + 6K^2Q^2 + K^4\right)\sinh K\sinh Q\Big] = 0 \end{aligned} \tag{26}$$

Note that parameters $p$ and $s$ may be expressed in terms of the viscous length scale

$$\ell_\nu = \left(\frac{\nu^2}{g}\right)^{\frac{1}{3}} \tag{27}$$

and the length scale related to surface tension

$$\ell_\sigma = \left(\frac{\sigma}{\rho g}\right)^{\frac{1}{2}} \tag{28}$$

Namely

$$p = \left(\frac{\ell_\nu}{h}\right)^3 \tag{29}$$

$$s = \left(\frac{\ell_\sigma}{h}\right)^2 \tag{30}$$

Given parameters $p$ and $s$, and scaled wave number $K$, the solution $Q$ of Eq. (26) yields the angular frequency (Eqs. (9) and (24) )

$$\omega = -i\frac{\nu}{h^2}\left(K^2 - Q^2\right) \tag{31}$$

For the ratio of the coefficients $A$ and $B$ we get (Eq. (19) )

$$\frac{B}{A} = \frac{K^2\cosh(K) - Q^2\cosh(Q)}{Q\left(K\sinh(K) - Q\sinh(Q)\right)} \tag{32}$$

Note that if $Q$ is a solution of Eq. (26), so is$-Q$. On the other hand, this sign does not matter when calculating $\omega$ or $f$ ( Eq. (15)). Henceforth we assume that the real part of $Q$ is positive, and thus $\tanh Q \to 1$ when $|Q| \to \infty$.

In the case of the small viscosity, where $p \to 0$ and $|Q| \to 1$, Eq. (26) can be solved approximately. This implies that in leading order Eq. (26) reduces to

$$(1 + sK^2)K\tanh K + p\,Q_0^4 = 0 \tag{33}$$

Or

$$Q_0^2 = -i\sqrt{\frac{(1 + sK^2)K\tanh K}{p}} \tag{34}$$

Here the negative sign has been chosen in order to get positive real part of angular frequency via Eq. (31). Further, according to the convention mentioned above, we have

$$Q_0 = \frac{1-i}{\sqrt{2}}\left(\frac{(1 + sK^2)K\tanh K}{p}\right)^{1/4} \tag{35}$$

A systematic expansion in terms of $p^{\frac{1}{4}}$ leads in the next two orders to

$$Q = Q_0 - \frac{K}{2\sinh(2K)} - \frac{K^2}{2Q_0}\frac{Y^2 + 6Y + 5}{Y(Y + 4)} \tag{36}$$

Where $Y = 4\sinh^2 K$. To this order we have for the angular frequency

$$\omega = \left[ \sqrt{\left(gk + \frac{\sigma k^3}{\rho}\right)\tanh(kh)} - \frac{\sqrt{2\nu k^2 \sqrt{\left(gk + \frac{\sigma k^3}{\rho}\right)\tanh(kh)}}}{2\sinh(2kh)} \right] - i \left[ \frac{\sqrt{2\nu k^2 \sqrt{\left(gk + \frac{\sigma k^3}{\rho}\right)\tanh(kh)}}}{2\sinh(2kh)} + 2\nu k^2 \frac{Y^2 + 5Y + 2}{Y(Y+4)} \right] \tag{37}$$

The first term of the real part is the well known dispersion relation of surface waves in ideal fluids. As for damping, the leading term is the first one in the second bracket, proportional to $\sqrt{\nu}$ , except in deep fluid. In deep fluid ($kh \to \infty$) this term vanishes and one gets the well know damping exponent $2\nu k^2$. The result (37) have first been published (for $\sigma = 0$) in [27] and then to higher orders in [3]. Note that taking into account surface tension is formally equivalent with replacing $p$ with $\frac{p}{(1+sK^2)}$ .

## 3. Wave modes

Plotting the real and imaginary parts of Eq. (26) on the complex $Q$ plane one usually observes several intersections, i.e. roots. They can be real, purely imaginary, or complex with nonzero real and imaginary parts. In the first two cases, the angular frequency (31) is purely imaginary. So these modes decay exponentially with time. Propagating modes are only possible if $Q$ is complex.

### 3.1 Modes with real $Q$

In this case, Eq. (31) implies that $Q < K$, since the decay rate cannot be negative. We have found numerically that at a given parameter settings ($s$, $p$ and $K$) there can exist zero, one or two real modes[1]. There is always a trivial solution $Q = K$. With this, however, we get from Eq. (15), $f = 0$. So this is solution irrelevant. In the case of nontrivial real solutions, the decay rate is always smaller than $\nu k^2$. Velocity components can be calculated from Eqs. (1), (2), (15) and (32). Provided that coefficient $A$ is real, $u$ is real, too, while $w$ is purely imaginary. As a result, there is a $90^o$ phase shift in their $x$-dependent oscillations. The $z$ dependence of the velocity components is shown in Figure 1.

### 3.2 Modes with imaginary $Q$

In this case the decay rate is always larger than $\nu k^2$. Such modes exist at any parameter setting. Moreover, there are infinitely many of them. Indeed, if $Q$ is purely imaginary and its modulus is large, Eq. (26) reduces to

$$p\left[Q^5 \cosh K \cosh Q - KQ^4 \sinh K \sinh Q\right] = 0 \tag{38}$$

substituting $Q = i\beta$

$$\beta \cosh K \cos \beta - K \sinh K \sin \beta = 0 \tag{39}$$

[1] Due to the symmetry of the solutions, we consider roots only in the first quadrant.

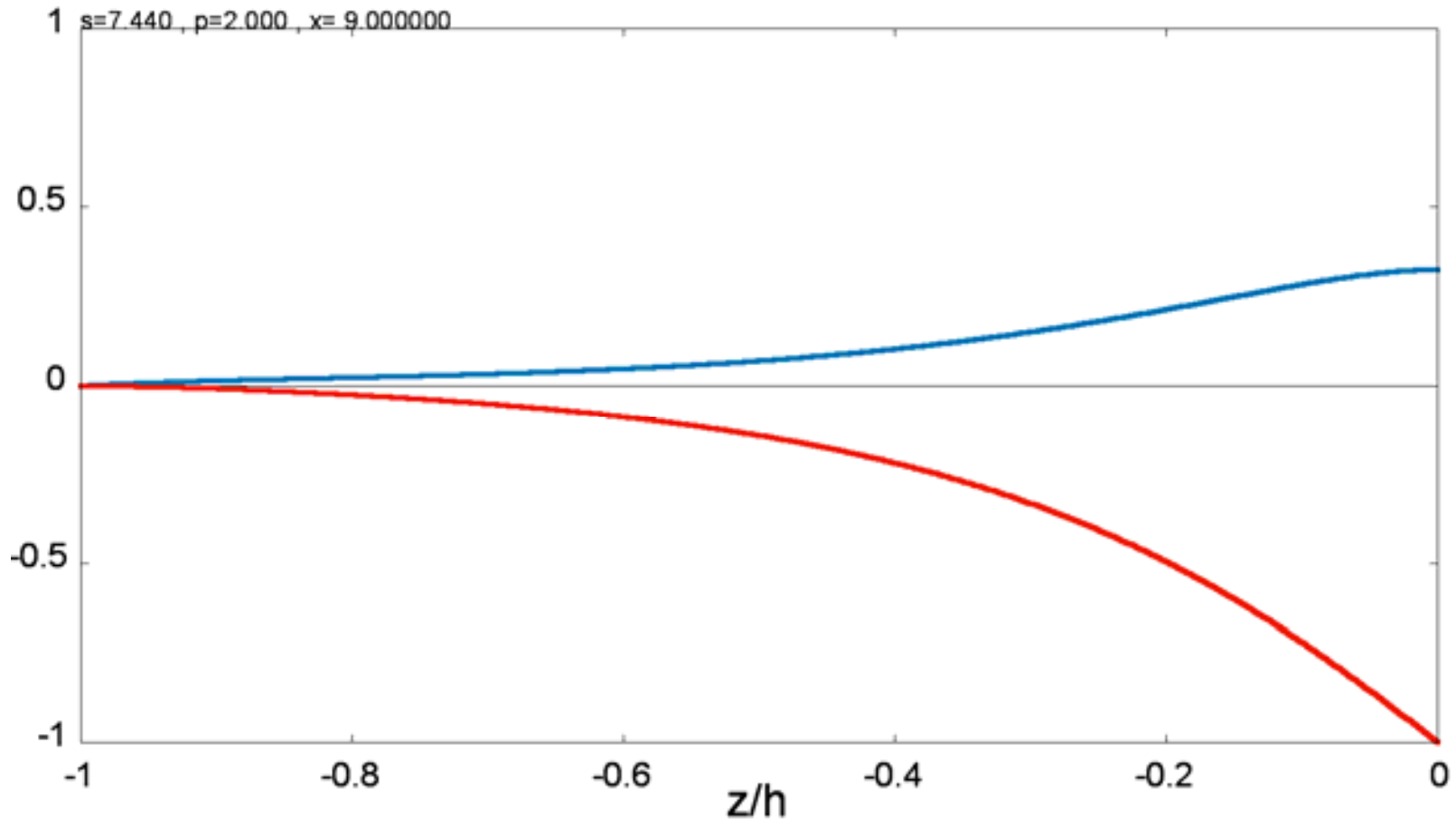

*Figure 1. z dependence of* $u$ *(blue) and* $w$ *(red) in case of real* $Q$*. Parameters: s=7.44, p=2.0 and K=1.0. Solution:* $Q = 4.35$*,* $\frac{z}{h} = -1$ *represents the bottom and* $\frac{z}{h} = 0$ *is the surface.*

Now, if $\beta = 2n\pi$ where $n$ being an integer, the left hand side is positive, and if $\beta = 2n\pi + \frac{\pi}{2}$, the left hand side is negative. Therefore, there is a root for any (arbitrarily large) $n$ between these values. The distance between imaginary roots is approximately constant and independent of viscosity. As before, the phase of velocity components does not change with depth, while there is a $90^o$ phase shift between the $x$ and $z$ components. As shown in Figure 2 , imaginary $Q$ causes an oscillatory behavior with depth.

### 3.3 Modes with complex $Q$

Based on our numerical investigations, we believe that at a given parameter setting at most one such mode can exist. If there exists one, then at the same parameter setting no mode with real $Q$ can exist. The phases of velocity components do change with depth. An oscillatory dependence on depth is in principle present, but much less pronounced than in the imaginary case. The amplitudes and phases are shown in Figure 3 and Figure 4.

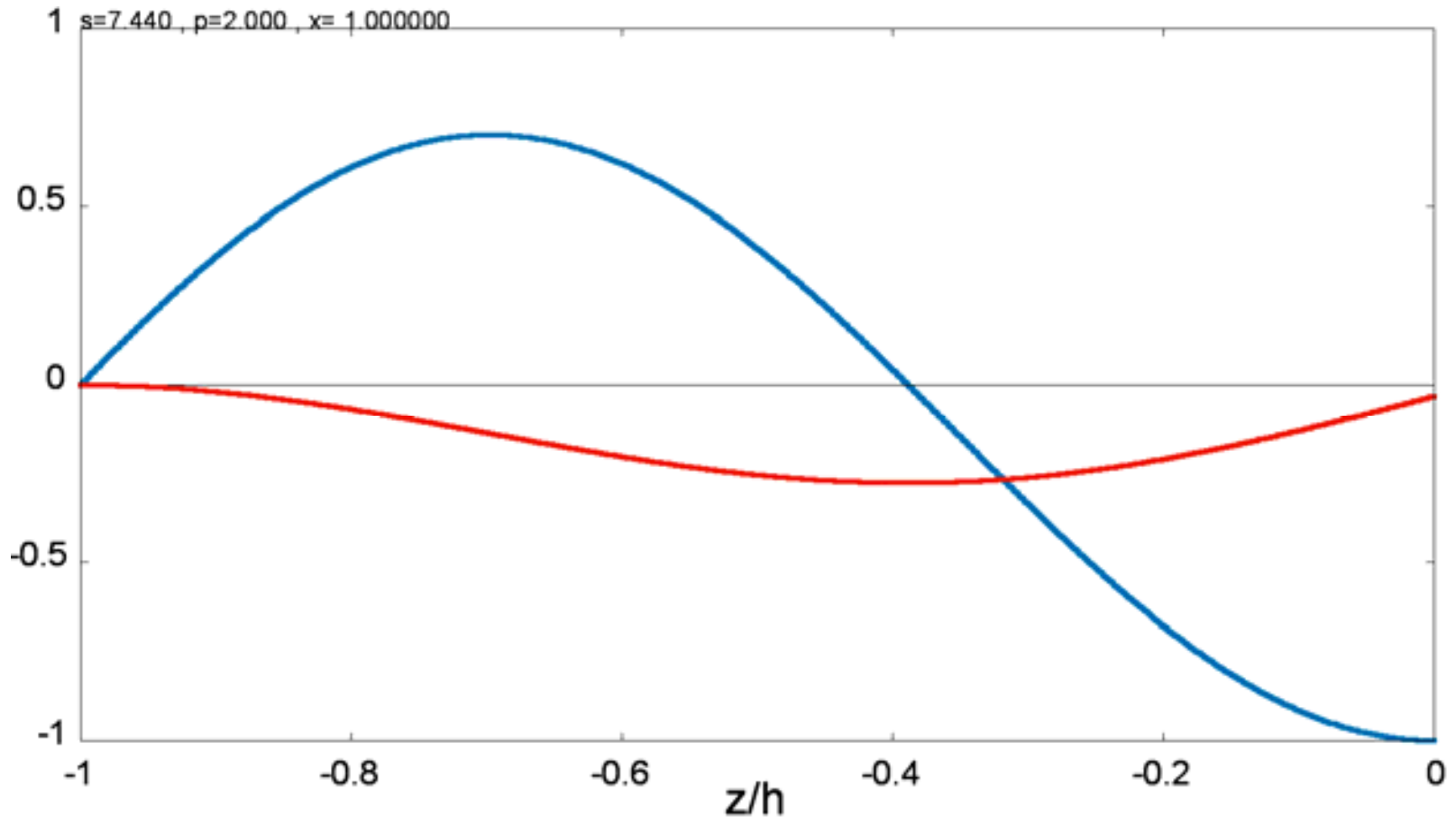

*Figure 2. $z$ dependence of $u$ (blue) and $w$ (red) in case of pure imaginary $Q$. Parameters: s=7.44, p=2.0 and K=9.0. Solution: $Q = 4.45\ i$.*

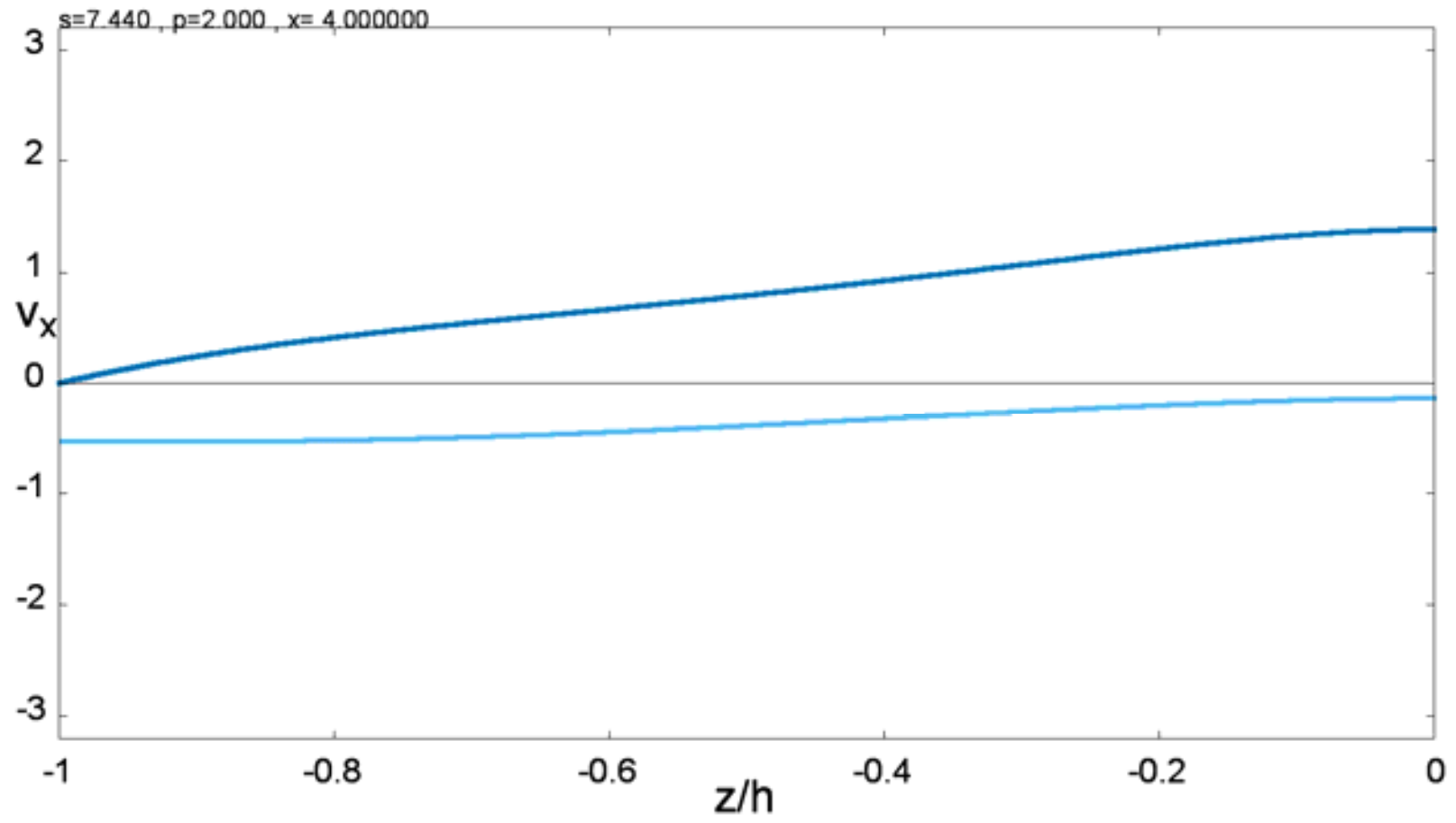

*Figure 3. $z$ dependence of $u$ in case of complex $Q$. Amplitude (blue) and phase (light blue) is shown. Parameters: s=7.44, p=2.0 and K=4.0. Solution: $Q = 2.66 + 1.09\ i$.*

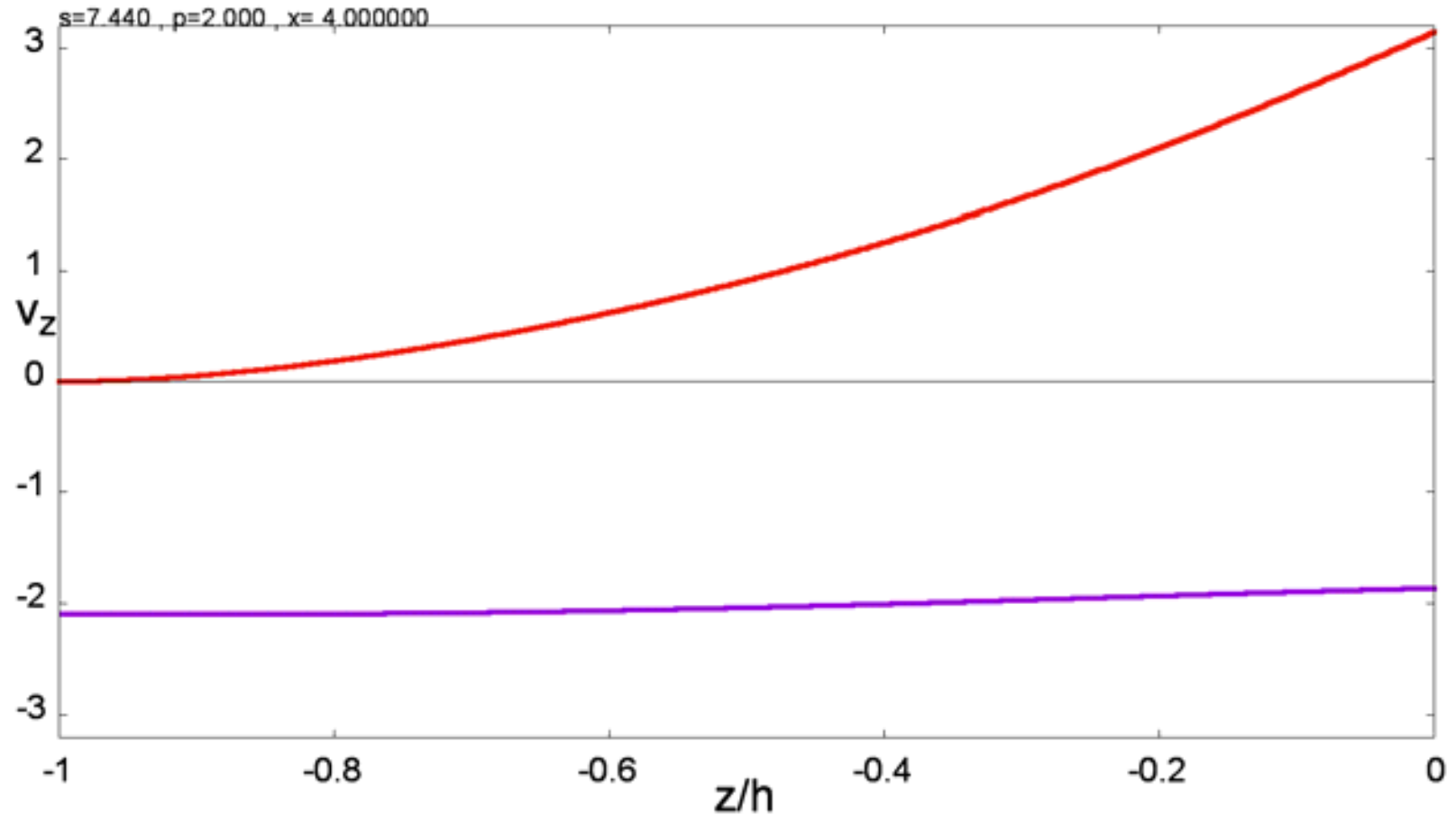

*Figure 4. $z$ dependence of $w$ in case of complex $Q$. Amplitude (red) and phase (purple) is shown. Parameters: s=7.44, p=2.0 and K=4.0. Solution: $Q = 2.66 + 1.09\ i$.*

## 4. Parameter dependence

If, at a given $s$ and $p$ parameters one adjusts $K$, the type and number of solutions changes. This is shown in Figure 5(a-i) where $s = 7.44$ and $= 2.0$ [2]. In Figure 5a one can see a single real (nontrivial) solution. This solution transforms through a bifurcation point illustrated in Figure 5b to an imaginary solution, shown in Figure 5c. As shown above, imaginary solutions always exist. This can be seen at larger scales in Figure 6a-6c.

Increasing the scaled wavenumber $K$ further, one can observe that two imaginary solutions collide (see.Figure 5d) and give rise to a complex solution (see Figure 5e). This complex solution gradually goes down to the real axis (Figure 5: e-h) and decays to two real solutions (Figure 5i) which survive any further increase of $K$.

Choosing the parameter values $s = 7.44$ and $p = 3.0$, one has a different scenario which is highlighted in Figure 7 (a-i). Here, only a single bifurcation takes place, namely, an imaginary solution and its mirror image collide at the origin (Figure 7f) and give rise to a second real solution (and its mirror image).

[2] For technical reasons Eq. (26) was divided by $K^2 Q \cosh(K) \cosh(Q)$ and the result was plotted.

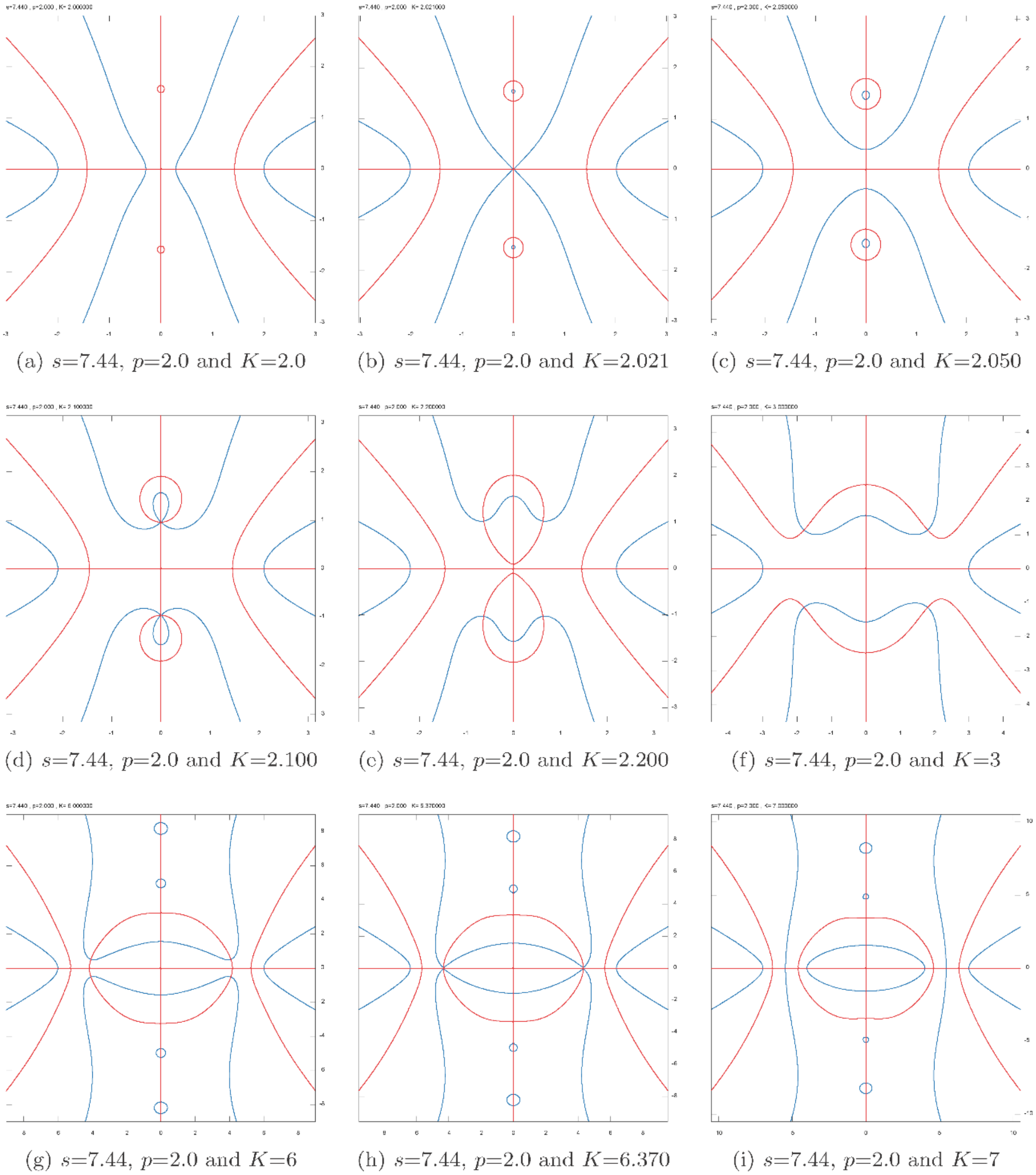

(a) $s$=7.44, $p$=2.0 and $K$=2.0
(b) $s$=7.44, $p$=2.0 and $K$=2.021
(c) $s$=7.44, $p$=2.0 and $K$=2.050
(d) $s$=7.44, $p$=2.0 and $K$=2.100
(e) $s$=7.44, $p$=2.0 and $K$=2.200
(f) $s$=7.44, $p$=2.0 and $K$=3
(g) $s$=7.44, $p$=2.0 and $K$=6
(h) $s$=7.44, $p$=2.0 and $K$=6.370
(i) $s$=7.44, $p$=2.0 and $K$=7

*Figure 5. Zero level lines of the real (blue) and imaginary (red) part of Eq.*(26) *plotted on the complex Q plane at parameter values p=2.0 and s=7.44.*

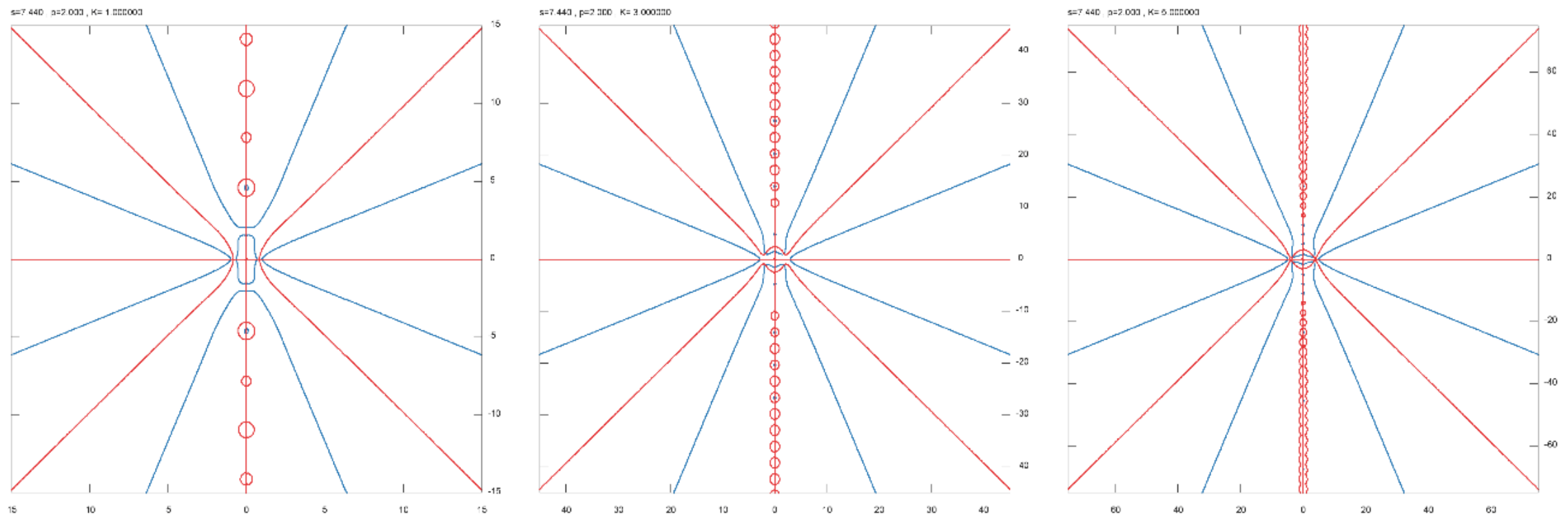

(a) $s$=7.44, $p$=2.0 and $K$=1.0 (b) $s$=7.44, $p$=2.0 and $K$=3.0 (c) $s$=7.44, $p$=2.0 and $K$=5.0

*Figure 6. Zero level lines of real and imaginary parts of Eq.* (26) *on the complex Q plane. Several solutions along the imaginary axis are shown. Note that they are roughly equidistant and independent of the parameters, but the scales of the figures are different.*

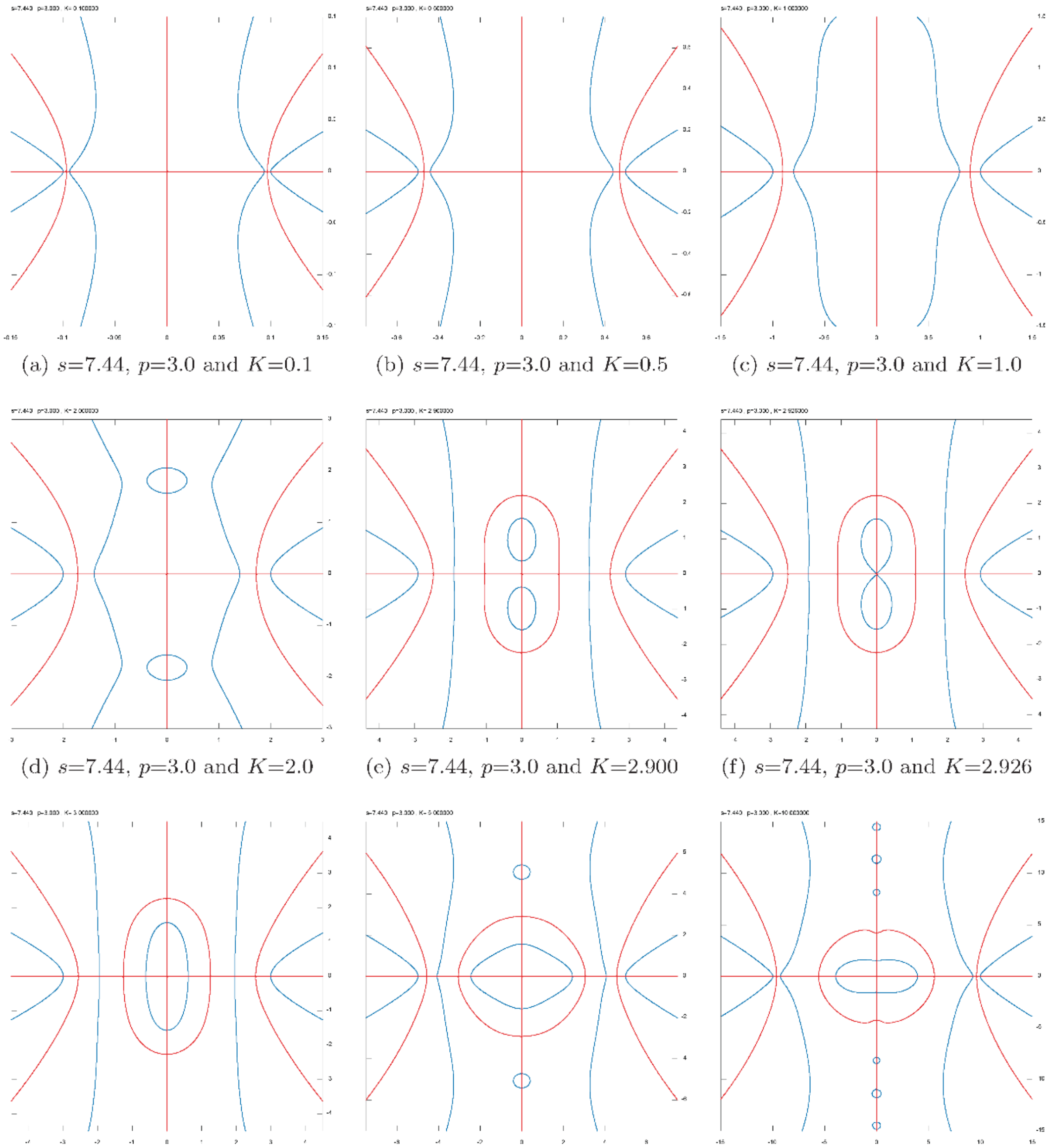

(a) $s$=7.44, $p$=3.0 and $K$=0.1 (b) $s$=7.44, $p$=3.0 and $K$=0.5 (c) $s$=7.44, $p$=3.0 and $K$=1.0

(d) $s$=7.44, $p$=3.0 and $K$=2.0 (e) $s$=7.44, $p$=3.0 and $K$=2.900 (f) $s$=7.44, $p$=3.0 and $K$=2.926

(g) $s$=7.44, $p$=3.0 and $K$=3 (h) $s$=7.44, $p$=3.0 and $K$=5 (i) $s$=7.44, $p$=3.0 and $K$=10

*Figure 7. Zero level lines of the real (blue) and imaginary (red) part of Eq.* (26) *plotted on the complex Q plane at parameter values p=3.0 and s=7.44.*

In all cases of creation of new solutions one can observe that approaching the critical parameter the zero level lines of the real part of Eq. (26) develop an edge, and at bifurcation they locally become two straight lines crossing each other on the real and/or the imaginary axis. In consequence, at the bifurcation point the derivative of the line is undetermined, i.e. has a form $\frac{0}{0}$. Note in passing that it is equivalent to the condition that the second derivative becomes infinite. Let us denote for brevity $\Re(Q)$ with $\alpha$ and $\Im(Q)$ with $\beta$, further, the left hand side of Eq. (26), divided by $K^2 Q\cosh(K)\cosh(Q)$ be $\xi(\alpha+i\beta)$. Then the zero level line of the real part of the $\xi$ is given by

$$\Re\left(\xi\left(\alpha+i\beta\right)\right)=0 \tag{40}$$

Taking its derivative with respect to $\alpha$, we get

$$\Re\left(\xi'\left(\alpha+i\beta\right)+i\xi'\left(\alpha+i\beta\right)\frac{d\beta}{d\alpha}\right)=0 \tag{41}$$

hence the derivative of the curve is written as

$$\frac{d\beta}{d\alpha}=\frac{\Re\left(\xi'\left(\alpha+i\beta\right)\right)}{\Im\left(\xi'\left(\alpha+i\beta\right)\right)} \tag{42}$$

It follows that at bifurcation

$$\xi'\left(\alpha+i\beta\right)=0 \tag{43}$$

must be satisfied, together with $\xi(\alpha+i\,\beta)=0$. Now it is easily seen that $\xi(\alpha+i\beta)$ is real along both the real and the imaginary axis, while $\xi'(\alpha+i\beta)$ is real along the real axis and imaginary along the imaginary axis. Therefore, in the case of a bifurcation on the real axis we have

$$\xi\left(\alpha\right)=0 \tag{44}$$

$$\xi'\left(\alpha\right)=0 \tag{45}$$

i.e., two real equations for the two real parameters $\alpha$ and $p$. Similarly, in the case of a bifurcation on the imaginary axis we have

$$\xi\left(i\beta\right)=0 \tag{46}$$

$$i\xi'\left(i\beta\right)=0 \tag{47}$$

again two real equations for two real parameters.
On the other hand, Eq. (26) shows that $\xi(Q)$ can be expressed as

$$\xi\left(Q\right)=\left(1+sK^2\right)F\left(Q\right)+pG\left(Q\right) \tag{48}$$

where

$$\mathrm{F}(Q)=\frac{\tanh K}{K}-\frac{\tanh Q}{Q} \tag{49}$$

$$\mathrm{G}(Q)=-4\frac{K^2+Q^2}{\cosh K\cosh Q}+\left(\frac{Q^4}{K^2}+2Q^2+5K^2\right)-\left(Q^4+6K^2Q^2+K^4\right)\frac{\tanh K}{K}\frac{\tanh Q}{Q} \tag{50}$$

In terms of these functions, we can write Eqs. (44) and (45) for bifurcations on the real and imaginary axis, respectively

$$F(\alpha)G'(\alpha)-G(\alpha)F'(\alpha)=0 \tag{51}$$

$$\frac{p}{1+sK^2}=-\frac{F(\alpha)}{G(\alpha)} \tag{52}$$

and

$$F(i\beta)G'(i\beta)-G(i\beta)F'(i\beta)=0 \tag{53}$$

$$\frac{p}{1+sK^2}=-\frac{F(i\beta)}{G(i\beta)} \tag{54}$$

In these cases, Eq. (51) or Eq (53) is solved numerically to get $\alpha$ or $\beta$, respectively, then the solution is inserted into Eqs. (52) and (54), respectively.
As for the transformation of an imaginary solution to a real one at the origin, mentioned above, for the critical $p(K)$ line one obtains

$$\frac{p}{1+sK^2}=-\frac{F(0)}{G(0)}=\frac{1-\dfrac{\tanh(K)}{K}}{5K^2-4\dfrac{K^2}{\cosh(K)}-K^3\tanh(K)} \tag{55}$$

Since both $F(Q)$ and $G(Q)$ are even functions of $Q$, it follows that $F'(0)=G'(0)=0$, hence Eqs. (51) and (53) are automatically satisfied for $\alpha=0$ and $\beta=0$, respectively. The results are plotted with and without considering surface tension in Figure 8. As shown in Figure 8, the red line means the onset of creation of complex solution at the imaginary line, i.e. the solution of Eqs. (53) and (54). In contrast, the blue line is corresponds to the solutions of Eqs. (51) and (52), while purple line is the onset of crossover from imaginary to real solution at the origin (Eq. (55)). It is obvious that these lines must have a common point. The lines partition the $K-p$ parameter space to four regions, denoted in the figures by Roman numbers:

I) Only imaginary solutions (infinitely many of them) are present.
II) There is a single real solution and there are infinitely many imaginary solutions.
III) There are two real solutions and there are infinitely many imaginary solutions.
IV) There is a single complex solution and there are infinitely many imaginary solutions.

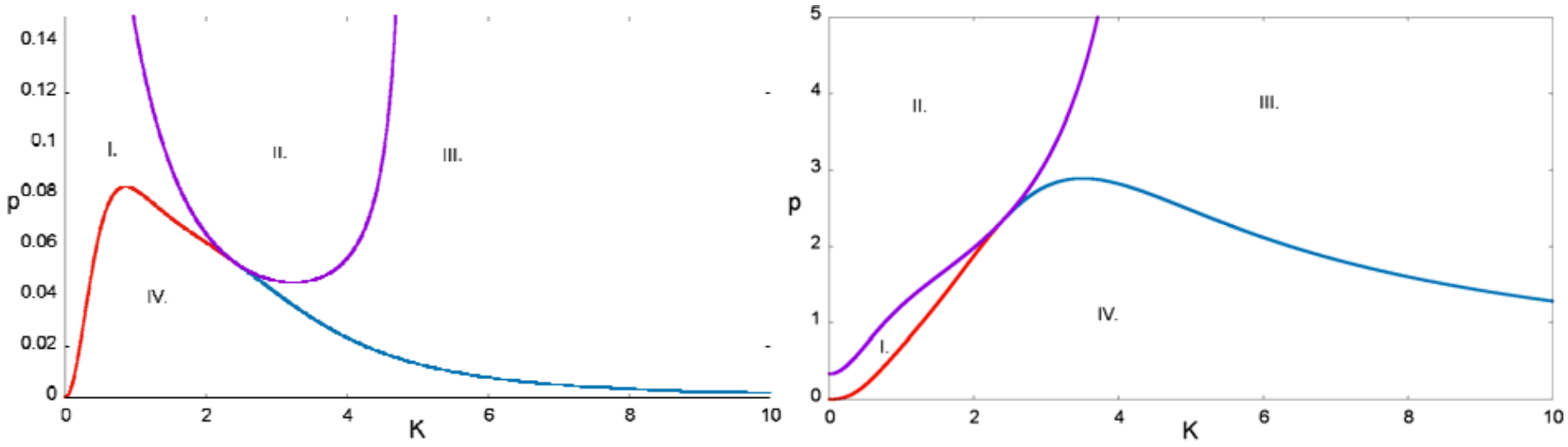

a) No surface tension assumed

b) with $s = 7.44$

*Figure 8. The maximal parameter $p$ versus the scaled wavenumber $K$.*

It is important to investigate asymptotics and some special cases of the bifurcation curves. To this end, we study the following cases.

1. Imaginary to complex $Q$ (red curve in Figure 8 a-b )

For large $K$ ($K \gg 1$), there is no solution. On the other hand, for small $K$ ($K \ll 1$) we have

$$Q = (1.1127 + 0.2509\, K^2) i \tag{56}$$

$$p = 0.53667\, K^2 \left(1 + (s - 3.8674) K^2\right) \tag{57}$$

2. Complex to real $Q$ (blue curve in Figure 8 a-b)

For small $K$, there is no solution, while for large $K$ we have

$$Q = 0.6823\, K \tag{58}$$

$$p = \frac{1.7200\,(1 + sK^2)}{K^3} \tag{59}$$

3. Imaginary to real $Q$ (purple curve in Figure 8 a-b )

For small $K$ we have

$$p = \frac{1}{3}\left[1 + \left(s - \frac{7}{5}\right) K^2\right] \tag{60}$$

At $K \approx 4.9435$ the parameter $p$ diverges as

$$p = \frac{0.8452 + 20.654\, s}{4.9435 - K} \tag{61}$$

4. The common point of the bifurcation parameter curves (red, blue and purple lines in Figure 8 a-b) satisfies

$$\frac{F(0)}{G(0)} = -\frac{p}{1 + sK^2} = \lim_{Q \to 0} \frac{F'(Q)}{G'(Q)} = \frac{F''(0)}{G''(0)} \tag{62}$$

since $F'(0) = G'(0) = 0$. This implies

$$F(0) G''(0) - G(0) F''(0) = 0 \tag{63}$$

The solution of this equation yields for the coordinates of the common point

$$K = 2.4152 \tag{64}$$

$$p = 0.05307\left(1 + 58332\, s\right). \tag{65}$$

## 5. Nonlinear waves

The fully nonlinear set of hydrodynamic equations gives rise to couplings among linear wave modes. As shown in the previous sections, at a given wavenumber $K$ there are infinitely many modes. Due to this, one cannot expect that a Korteweg-de Vries-type (henceforth KdV) differential equation for the wave amplitude takes into account all the features of a general solution. Possible simplifications are due to special circumstances. E.g. if $\frac{a}{h} \sim \sqrt{\frac{h}{l}}$ then the viscous KdV equation becomes applicable [1]. Here we speculate if a strategy is applicable for viscous nonlinear waves similar to that in the case of ideal fluids to derive the KdV equation. Indeed, if it is sufficient to consider the first few terms of the Taylor expansion in terms of the vertical coordinate $z$ (measured from the flat bottom, note for the different convention used in Section 2.), one may derive evolution equations for them from the Navier-Stokes equations. Particularly, we look for the solution in the form

$$u = \sum_{j=1}^{n} u^{(j)}(x,t)\left(\frac{z}{h}\right)^{j} \tag{66}$$

$$w = -\sum_{j=2}^{n+1} \left(\frac{h}{j}\right) u_x^{(j-1)}\left(x,t\right)\left(\frac{z}{h}\right)^{j} \tag{67}$$

which automatically satisfies the incompressibility condition $\nabla \mathbf{v} = 0$. In the linear case, the rotation of the Navier-Stokes equation gives (Eq. (5)) for $n = 4$

$$u_t^{(1)} = 2\nu u_{xx}^{(1)} - 6\frac{\nu}{h^2} u^{(3)} \tag{68}$$

$$u_t^{(2)} = 2\nu u_{xx}^{(2)} - \frac{1}{2} u^{(1)} u_x^{(1)} + 12\frac{\nu}{h^2} u^{(4)} \tag{69}$$

by setting to zero the coefficients of $z^0$ and $z^1$. Note that the coefficients of the higher powers of $z$ would contain higher than fourth order terms in Eq. (66). In general we have $n - 2$ equations for the coefficient functions $u^j$. For the remaining two coefficients and the surface position $\eta(x,t)$ we have the boundary conditions

$$w = \eta_t + u\eta_x \tag{70}$$

$$\left(1 - \eta_x^2\right)\left(u_z + w_x\right) - 4\eta_x u_x = 0 \tag{71}$$

$$\frac{p_0 - p}{\rho} = 2\nu \frac{\eta_x}{1 + \eta_x^2}\left(u_z + w_x\right) \tag{72}$$

where $z = h + \eta$. Eqs. (71) and (72) express the continuity of the tangential and the normal component of the strain at the surface, respectively. $p_o$ in Eq. (72) stands for the atmospheric pressure at the surface (considered a constant). Pressure $p$ is obtained by integrating the vertical components of the NS equation

$$\frac{p}{\rho}=\frac{B}{\rho}-gz+\int_0^z dz\left(\nu\,\Delta w-w_t-u\,w_x-w\,w_z\right) \tag{73}$$

Here $B=B(x,t)$ stands for the bottom pressure, which satisfies (by taking the $x$ derivative of Eq. (73) )

$$B_x=\rho\nu u_{zz}\left(z=0\right)=\frac{2\rho\nu}{h^2}u^{(2)} \tag{74}$$

Eventually, by inserting the pressure (73) into Eq. (72) we have four equations, Eqs. (70)-(74) for the quantities $\eta$, $B$, $u^{(n-1)}$, $u^{(n)}$. Explicitly, for $n=4$ we have beyond Eqs. (68), (69) and (74)

$$\bar{\eta}_t+\left[\frac{1}{2}u^{(1)}\bar{\eta}^2+\frac{1}{3}u^{(2)}\bar{\eta}^3+\frac{1}{4}u^{(3)}\bar{\eta}^4+\frac{1}{5}u^{(4)}\bar{\eta}^5\right]_x=0 \tag{75}$$

$$\begin{aligned}&u^{(1)}+2u^{(2)}\bar{\eta}+3u^{(3)}\bar{\eta}^2+4u^{(4)}\bar{\eta}^3-\frac{h^2}{2}u^{(1)}_{xx}\bar{\eta}^2-\frac{h^2}{3}u^{(2)}_{xx}\bar{\eta}^3-\frac{h^2}{4}u^{(3)}_{xx}\bar{\eta}^4-\frac{h^2}{5}u^{(4)}_{xx}\bar{\eta}^5\\&-\frac{4h^2\bar{\eta}_x}{1-h^2\bar{\eta}_x^2}\left[u^{(1)}_x\bar{\eta}+u^{(2)}_x\bar{\eta}^2+u^{(3)}_x\bar{\eta}^3+u^{(4)}_x\bar{\eta}^4\right]=0\end{aligned} \tag{76}$$

$$\begin{aligned}&\frac{B-p_0}{\rho h}-g\bar{\eta}+\frac{h}{6}u^{(1)}_{xt}\bar{\eta}^3+\frac{\nu}{2}u^{(1)}_{xx}\bar{\eta}^2-\frac{\nu}{h}u^{(1)}_x\bar{\eta}\\&+\frac{h}{12}u^{(2)}_{xt}\bar{\eta}^4+\frac{\nu}{3}u^{(2)}_{xx}\bar{\eta}^3-\frac{\nu}{h}u^{(2)}_x\bar{\eta}^2+\frac{h}{8}\left[u^{(1)}u^{(1)}_{xx}\bar{\eta}^4-\left(u^{(1)}_x\right)^2\right]\bar{\eta}^4\\&+\frac{h}{20}u^{(3)}_{xt}\bar{\eta}^5+\frac{\nu}{4}u^{(3)}_{xx}\bar{\eta}^4-\frac{\nu}{h}u^{(3)}_x\bar{\eta}^3+\left[\frac{h}{15}u^{(1)}u^{(2)}_{xx}+\frac{h}{10}u^{(2)}u^{(1)}_{xx}-\frac{h}{6}u^{(1)}_xu^{(2)}_x\right]\bar{\eta}^5\\&+\frac{h}{30}u^{(3)}_{xt}\bar{\eta}^6+\frac{\nu}{5}u^{(3)}_{xx}\bar{\eta}^5-\frac{\nu}{h}u^{(3)}_x\bar{\eta}^4\\&+\left[\frac{h}{18}u^{(2)}u^{(2)}_{xx}+\frac{h}{24}u^{(1)}u^{(3)}_{xx}-\frac{h}{18}\left(u^{(2)}_x\right)^2+\frac{h}{10}u^{(2)}u^{(1)}_{xx}-\frac{h}{8}u^{(1)}_xu^{(3)}_x+\frac{h}{12}u^{(3)}u^{(1)}_{xx}\right]\bar{\eta}^6\\&+\left[\frac{h}{28}u^{(2)}u^{(3)}_{xx}+\frac{h}{21}u^{(3)}u^{(2)}_{xx}+\frac{h}{35}u^{(1)}u^{(4)}_{xx}+\frac{h}{14}u^{(4)}u^{(1)}_{xx}-\frac{h}{12}u^{(2)}_xu^{(3)}_x-\frac{h}{10}u^{(1)}_xu^{(4)}_x\right]\bar{\eta}^7\\&+\left[\frac{h}{40}u^{(2)}u^{(4)}_{xx}+\frac{h}{32}u^{(3)}u^{(3)}_{xx}+\frac{h}{24}u^{(4)}u^{(2)}_{xx}-\frac{h}{32}\left(u^{(3)}_x\right)^2-\frac{h}{15}u^{(2)}_xu^{(4)}_x\right]\bar{\eta}^8\\&+\left[\frac{h}{45}u^{(3)}u^{(4)}_{xx}+\frac{h}{36}u^{(4)}u^{(3)}_{xx}-\frac{h}{20}u^{(3)}_xu^{(4)}_x\right]\bar{\eta}^9+\frac{h}{50}\left[u^{(4)}u^{(4)}_{xx}-\left(u^{(4)}_x\right)^2\right]\bar{\eta}^{10}\\&+\frac{8\nu h\bar{\eta}_x^2}{1-h^4\bar{\eta}_x^4}\left[u^{(1)}_x\bar{\eta}+u^{(2)}_x\bar{\eta}^2+u^{(3)}_x\bar{\eta}^3+u^{(4)}_x\bar{\eta}^4\right]=0\end{aligned} \tag{77}$$

Where

$$\bar{\eta}=1+\frac{\eta}{h} \tag{78}$$

## 6. Discussion and summary

### 6.1 Comparing dispersion relation (26) with previous results

Our Eq. (26) is the same as Eq. (13) of [3], when surface tension is not considered. By introducing

$$\varepsilon = \left( \frac{4\nu^2 k^3}{g} \right)^{\frac{1}{4}} = \left(4pK^3\right)^{\frac{1}{4}} \tag{79}$$

$$M = Q \qquad \sigma = \omega \tag{80}$$

Hunt reported the following characteristic equation

$$\begin{aligned} &\left(\frac{Q}{K}\tanh K - \tanh Q\right)\cosh K\ \cosh Q + \frac{\varepsilon^4}{4}\left[-4\frac{Q}{K}\left(1+Q^2\right)\right. \\ &\left. + \frac{Q}{K}\left(\left(\frac{Q}{K}\right)^4 + 2\left(\frac{Q}{K}\right)^2 + 5\right)\cosh K \cosh Q - \left(\left(\frac{Q}{K}\right)^4 + 6\left(\frac{Q}{K}\right)^2 + 1\right)\sinh K \sinh Q\right] = 0 \end{aligned} \tag{81}$$

In this case one can rewrite $Q$ in terms of $K$ and $\epsilon$

$$2K^2 S = i\varepsilon^2\left(Q^2 - K^2\right) \tag{82}$$

where $S = \frac{\omega}{\sqrt{gk}}$. He expanded the Eq. (81) in terms of different powers of $\epsilon$ and discussed both deep water and shallow water limits. He also proved the identity of Biesel's results [27] with his. By non-dimentionalizing, Meur represent the same relation as Eq. (26) if $\sinh(\mu)$ is changed to $\cosh(\mu)$ in the last term of Eq (13) in [1]. To get dimensionless fields and variables he choosed a characteristic horizontal length $l$, which is the wavelength, a characteristic vertical length $h$ which is the water's height, and the amplitude $A$ of the propagating perturbations.

$$c_o = \sqrt{gh} \qquad \alpha = \frac{A}{h} \qquad \beta = \frac{h^2}{l^2} \qquad \mathrm{Re} = \frac{c_o h}{\nu} \tag{83}$$

Comparing with our notation, one can find

$$\mu = Q \qquad k = \frac{K}{\sqrt{\beta}} \qquad \mathrm{Re} = \sqrt{\frac{1}{p}} \tag{84}$$

Sanochkin studied the damping of gravitational–capillary waves [28]. He analyzed the effect of viscosity on free-surface waves without any restrictions on the values of the viscosity coefficient and wavelength. In his notation

$$s = -i\omega,\ \ a = K,\ \ b = \frac{1}{2}\varepsilon^2,\ \ w = -iS,\ \ q = \frac{w}{b} = \frac{Q^2}{K^2} - 1 \tag{85}$$

His dispersion relation deffers from ours. If, however, $\cosh\left(a\sqrt{q}\right)$ is changed to $\cosh\left(a\sqrt{1+q}\right)$ and $\sinh\left(\frac{a}{\sqrt{1+q}}\right)$ is changed to $\sinh\left(a\sqrt{1+q}\right)$, then Eq. (1.5) in [28] coincides with our dispersion relation (26).

### 6.2 No surface waves exist in shallow fluid layers

We considered possible linear surface waves and their parameter dependence in a viscous, incompressible open surface fluid layer with surface tension. Viscosity not only damps waves, but it can even prevent their propagation [28]. Mathematically, propagation which corresponds to the real part of the complex angular frequency, appears only in region IV (Figure 8). If surface tension were negligible (Figure 8a), we can state that

1. No gravity waves can propagate if $p > 0.085$ [3]
2. Even if $p < 0.085$, neither very long, nor very short waves can propagate.

Certainly, in such shallow fluid layers surface tension may not be negligible. If we take surface tension into account, we get Figure 8b. If one plots the maxima of these curves versus $s$, one obtains Figure 9. The curve starts linearly with a small slope (see inset). For large $s$ values we again have a linear dependence with a much larger slope.

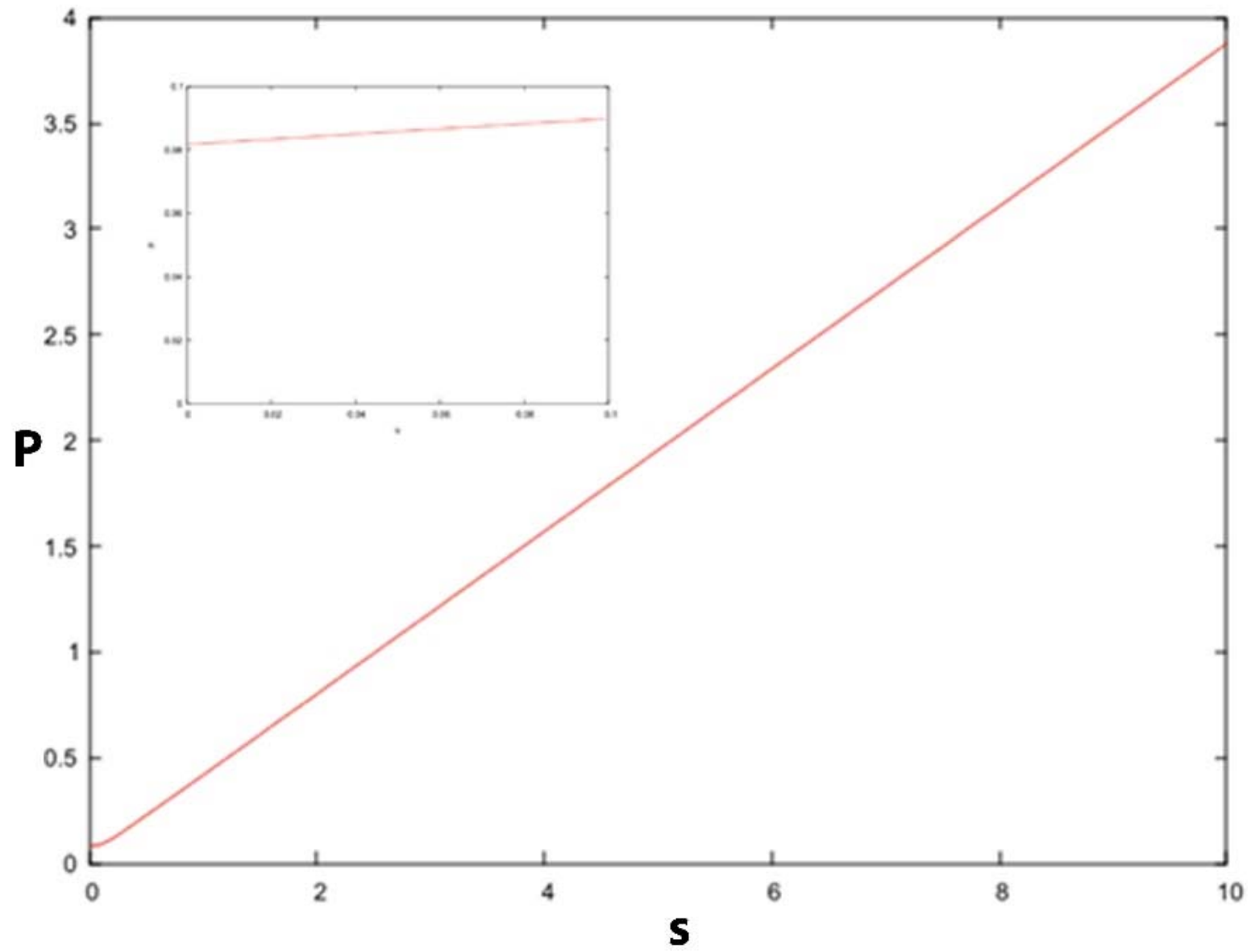

*Figure 9. Maximal parameter p versus scaled surface tension s. For parameters below the curve wave propagation is possible. Inset: blowup of region near the origin.*

The linear dependence at both small and large values of surface tension follows directly from the fact for large $s$ values, we have a linear dependence, namely $p \approx 0.384\, s$. This implies that wave propagation is not possible if

$$h < 2.6 \times \frac{\rho v^2}{\sigma} \tag{86}$$

This gives the limit for water $h < 3.56 \times 10^{-8}\, m$. As for glycerine, its parameters correspond to the beginning of the curve at $s = 0.086$. So, the limit obtained for gravity waves is not significantly changed. In summary, no surface waves can propagate in a fluid layer that is thin enough. However, this may be checked in case of glycerine, where this critical layer thickness is $7.7\, mm$.

[3] In the case of water, this translates to $h < 0.1\, mm$, and for glycerine to $h < 7.7\, mm$.

### 6.3 Limitations of the nonlinear approximation

We have to be aware that it never happens that both $K$ and $Q$ are small for the linear modes. Indeed, if we plot the modulus of $Q$ on the parameter plane, we get Figure 10. One can see that there is ridge in the middle of the picture (roughly vertically). It is now a question how it behaves near the lower left corner. To this end, we plotted the modulus of $Q$ along the critical $p(K)$ curve. The result is shown in Figure 11.

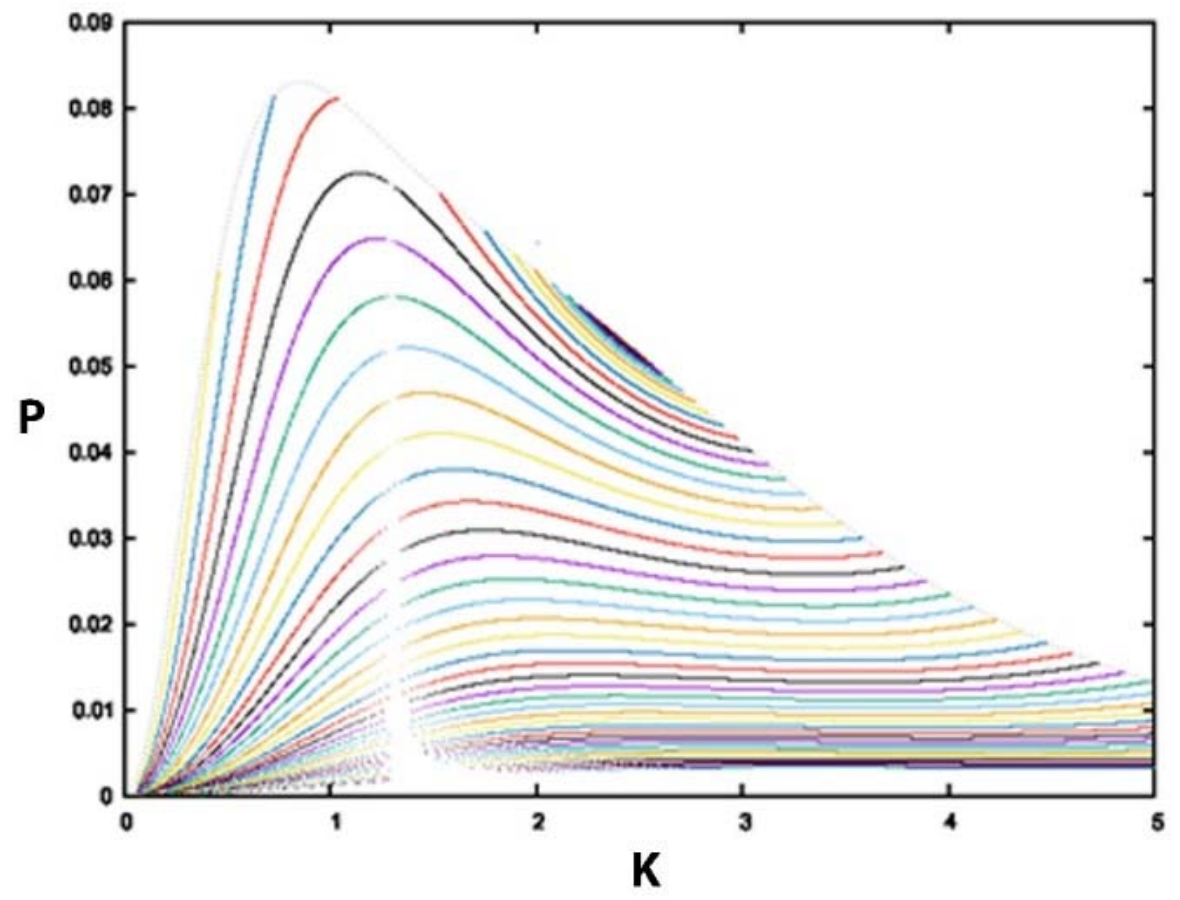

*Figure 10. Level hights of modulus of Q in case of zero surface tension.*

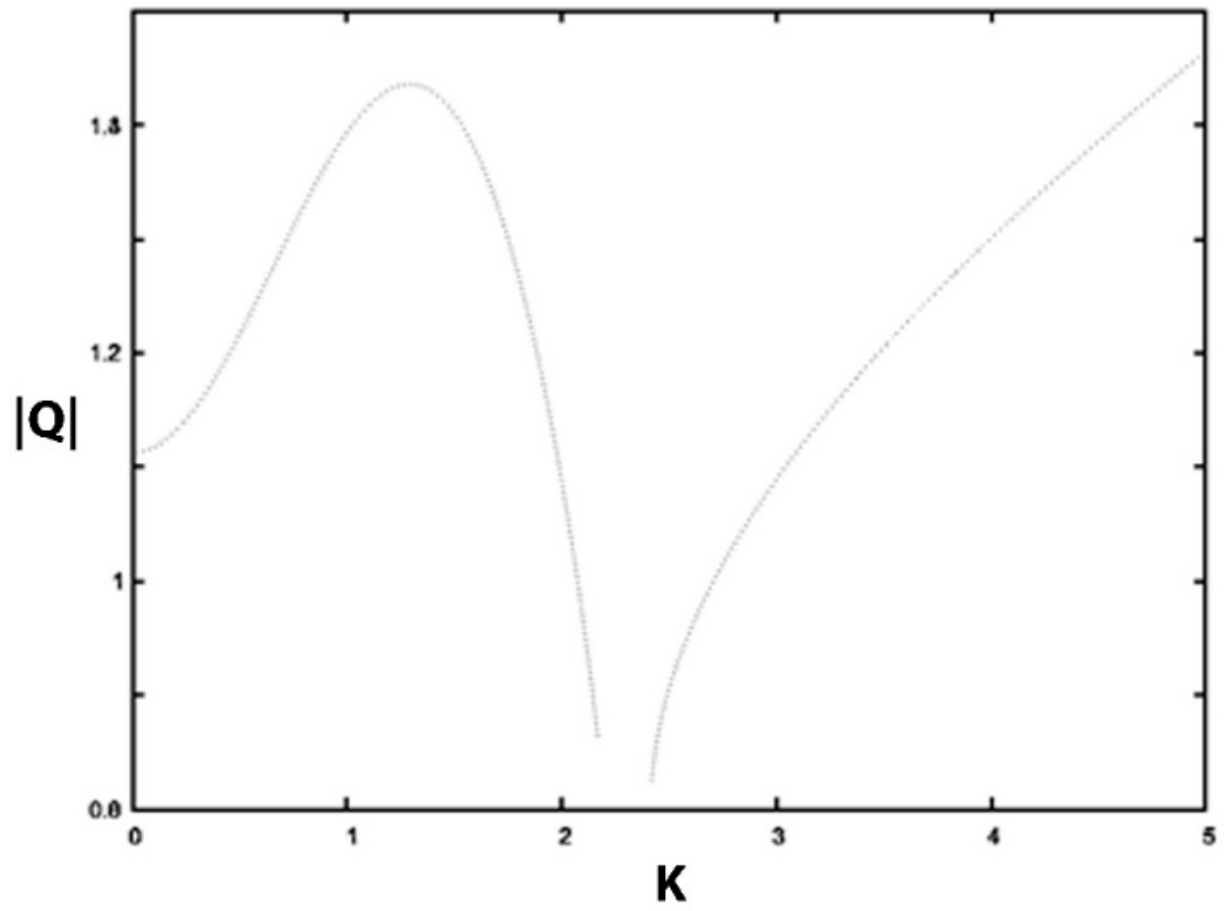

*Figure 11. Modulus of Q along the critical curve.*

According to Figure 11, where $K$ is small, $Q$ remains finite (cf. (56)). This means that even for very small $K$, a finite number of terms in Taylor series (66) and (67) leads to a finite error which may only be decreased with increasing $n$. Note also that we tacitly assumed that the coupling to the nonpropagating modes with large $Q$ values was negligible. This condition is fullfilled if the amplitude is small or $\frac{a}{h} \ll 1$.